\documentclass[twocolumn]{aastex631}
\newcommand{\lya}{Ly$\alpha$}
\newcommand{\lyb}{Ly$\beta$}
\newcommand{\ha}{H$\alpha$}
\newcommand{\hb}{H$\beta$}
\newcommand{\hg}{H$\gamma$}
\newcommand{\ca}{\ion{Ca}{2}}
\newcommand{\cah}{\ion{Ca}{2}~H}
\newcommand{\cak}{\ion{Ca}{2}~K}
\newcommand{\cahk}{\ion{Ca}{2}~H and K}
\newcommand{\cawav}{\ion{Ca}{2}~8542\,\AA}
\newcommand{\hed}{\ion{He}{1}~D3}
\newcommand{\heir}{\ion{He}{1}~10830\,\AA}
\newcommand{\oxy}{\ion{O}{1}~7772\,\AA}
\newcommand{\oo}{\ion{O}{1}}
\newcommand{\fe}{\ion{Fe}{1}~6302\,\AA}
\newcommand{\he}{\ion{He}{1}}

\newcommand{\mghk}{\ion{Mg}{2}~h and k}

\newcommand{\kms}{km\,s$^{-1}$}

\newcommand{\WHz}{W\,m$^{-2}$\,sr$^{-1}$\,Hz$^{-1}$}
\newcommand{\vmic}{v_{\rm mic}}
\newcommand{\RN}[1]{\textup{\lowercase\expandafter{\romannumeral#1}}}

\usepackage{soul}

\submitjournal{ApJ}

\shorttitle{Dark off-limb gap}
\shortauthors{Kuridze et al.}

\begin{document}

\title{Dark off-limb gap: manifestation of temperature minimum and dynamic nature of the chromosphere}


\correspondingauthor{D. Kuridze}
\email{dak21@aber.ac.uk}

\author[0000-0003-2760-2311]{David Kuridze} 
\affiliation{Department of Physics, Aberystwyth University, Ceredigion, SY23 3BZ, UK}
\affiliation{Abastumani Astrophysical Observatory, Mount Kanobili, 0301, Abastumani, Georgia}

\author[0000-0002-5778-2600]{Petr Heinzel}
\affiliation{Astronomical Institute, The Czech Academy of Sciences, 25165 Ond\v{r}ejov, Czech Republic}
\affiliation{University of Wroclaw, Centre of Scientific Excellence - Solar and Stellar Activity, Kopernika 11, 51-622 Wroclaw, Poland}

\author[0000-0002-7444-7046]{J\'{u}lius Koza} 
\affil{Astronomical Institute, Slovak Academy of Sciences, 059 60 Tatransk\'{a} Lomnica, Slovakia}

\author[0000-0003-4162-7240]{Ramon Oliver}
\affiliation{Departament de F\'{\i}sica, Universitat de les Illes Balears, E-07122 Palma de Mallorca, Spain}
\affiliation{Institute of Applied Computing \& Community Code (IAC3), UIB, Spain}


\begin{abstract}
We study off-limb emission of the lower solar atmosphere using
high-resolution imaging spectroscopy in the \hb\ and \cawav\ lines
obtained with the CHROMospheric Imaging Spectrometer (CHROMIS) and the
CRisp Imaging SpectroPolarimeter (CRISP) on the Swedish 1-m Solar
Telescope. The \hb\ line wing images show the dark intensity gap
between the photospheric limb and chromosphere which is absent in
  the \ca\ images. We calculate synthetic spectra of the off-limb
emissions with the RH code in the one-dimension spherical geometry and
find good agreement with the observations.  The analysis of synthetic
line profiles shows that the gap in the \hb\ line wing images maps the
temperature minimum region between the photosphere and chromosphere
due to the well known opacity and emissivity gap of Balmer lines in
this layer.  However, observed gap is detected farther from the line
core in the outer line wing positions than in the synthetic profiles.
We found that an increased microturbulence in the model chromosphere
is needed to reproduce the dark gap in the outer line wing, suggesting
that observed \hb\ gap is the manifestation of the temperature minimum
and the dynamic nature of the solar chromosphere.  The temperature
minimum produces a small enhancement 
in synthetic \ca\ line-wing intensities. Observed off-limb \ca\ line-wing
emissions show similar enhancement below temperature minimum layer
near the edge of the photospheric limb.
\end{abstract}



\section{Introduction}
\label{intro}

The solar atmosphere under non-ideal MHD approximation evolves as a
magnetized fluid that is governed by equations of hydrodynamics
representing conservation of mass, momentum, and energy. These are
coupled with radiative transfer equation and induction equation for
magnetic field. Within the three-dimensional radiation-MHD (3D R-MHD)
modeling, basic parameters of the solar atmosphere such as
temperature, velocity, density, and magnetic field result from solving
them in spatial and temporal domain. However, the 3D R-MHD modeling
has limitations, such as ad-hoc replacing the viscosity and resistance
by numerical terms that are independent of temperature
\citep{Leenaarts2020}. One of the most puzzling aspects of the solar
atmosphere is breaking the radiative equilibrium in the layers above
the photosphere. Therefore deposition of non-thermal energy and
conversion into heat must occur in the chromosphere and corona---thus
they are hotter than the radiative equilibrium assumes. The question
``How much hotter?'' was answered by analysis of ground and space
spectroscopic data acquired by various techniques.

Early solar physics utilized solar eclipses as the primary source of
quantitative chromospheric data \citep[see][and references
  therein]{Judgeetal2019}. Time series of flash spectra, acquired with
high cadence in brief moments after and before the second and third
contact, respectively, yielded first hints on physical conditions in
the chromosphere. For example, an extensive set of the high-cadence
chromospheric flash spectra, obtained at the 1952 eclipse in Khartoum,
was analyzed in \citet{Athayetal1954}. Recently, high-cadence flash
spectra ($\geq 8$\,ms cadence), acquired by a novel experiment during
the 2017 August 21 eclipse, were studied in \citet{Judgeetal2019}.

Lack of detailed knowledge of heating mechanisms in the layers, where
radiative equilibrium breaks down, forced to construct one-dimensional
(1D) semi-empirical (SE) models where the temperature structure of the
atmosphere (i.e., $T$ versus the column mass or height) is inferred
from long-slit spectroscopy by the radiative transfer calculations
assuming non-local thermodynamic equilibrium (non-LTE; i.e.,
departures from LTE). The non-LTE problem takes consistently into
account weak coupling of radiation field with the matter in the
chromosphere. The 1D SE models are primarily based on on-disk
observations of continua and several strong spectral
lines. Temperature is the variable that most strongly controls the
continuum and spectral line shapes and intensities yielded by the
classical SE models
\citep{Gingerich1971,Vernazzaetal1981,Fontenlaetal1993,RuttenandUitenbroek2012,HeinzelandStepan2019}. This
sensitivity arises from the exponential and power dependence with $T$
in the excitation-ionization processes \citep{Gray2008}. Their
characteristic feature is a distinct temperature minimum (TM)
separating the upper photosphere, dominated by the gas pressure, from
the lower chromosphere where the magnetic field gains a dominance over
the gas.
Unfortunately, many underlying processes, such as granulation, waves, flows,
and jets, are not reproduced by 1D SE models as they are 
obtained using the hydrostatic equilibrium and 
plane-parallel geometry approximation \citep{Rutten2002}. 
These processes are better
represented by computationally much more expensive 3D-MHD models which
also have limitations mentioned in the first paragraph of this
section. This may justify an employment of 1D modeling particularly
when the challenging off-limb viewing geometry is considered.

\begin{figure*}
\includegraphics[width=\textwidth]{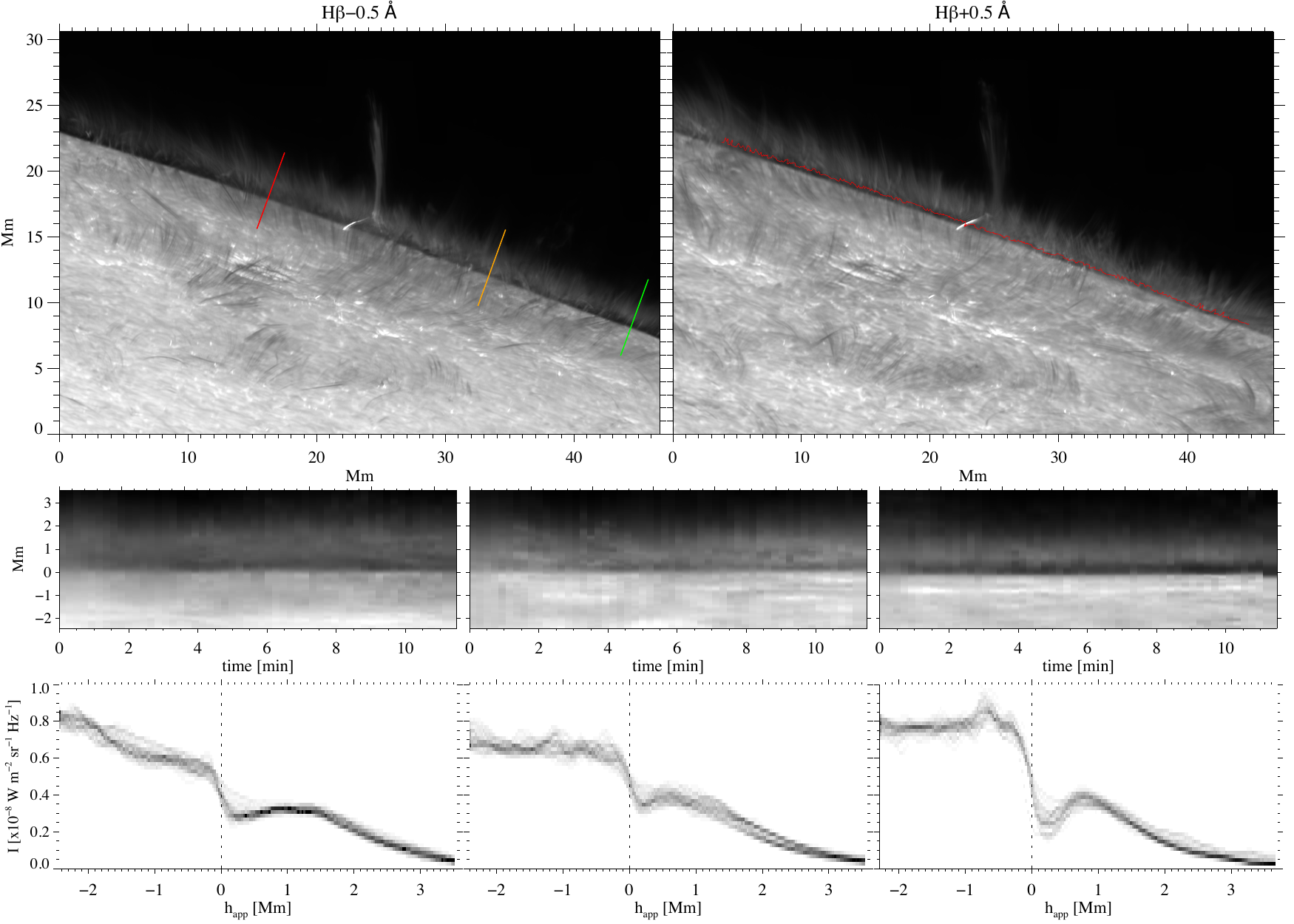}
\caption{Top: CHROMIS \hb\ wing images at
  $\Delta\lambda=-0.5$\,\AA\ (left) and $+0.5$\,\AA\ (right). The red
  contour in the top right panel follows the locations of minimum
  intensity in the dark gap. Middle: Time-distance diagrams of
  intensities extracted along the dashes in the top left
  panel. Bottom: Superimposition of the extracted intensities at
  individual times shown in the middle panels. The zero apparent
  height of the photospheric limb corresponds to the position of
  inflection point (the vertical dashed line) of the radial intensity
  gradients along the dashes pertinent to the \hb\ far wing image at
  $\Delta\lambda =-1.2$\,\AA.}
\label{fig1}
\end{figure*}
    
It has been shown that specific monochromatic intensities of some
chromospheric lines sample the TM layer and can be used to study the
location and properties of this layer. For example, the K$_1$ feature
of \cak\ line can map the TM in the solar and stellar atmospheres
\citep[e.g.,][]{Shine1975,AyresandLinsky1976,Avrett1985,Mauas2000}. The
most cited paper on SE non-LTE models by \citet{Vernazzaetal1981}
places the TM with the local temperature $\sim 4200$\,K about 500\,km
above the visible solar surface corresponding to the optical depth
unity at 5000\,\AA. Since then the temperature and the location of TM
have been a subject of an intensive
discussion. \citet{Fontenlaetal2007} list previous papers preferring
warm or cooler TM models, divided further as a single component or
bifurcated, and present a single-component model, which matches
several apparently contradictory observations. For example, SE models
of sunspot by \citet{Maltby1986} place TM at the height of around
$300-550$\,km. Analyzing the flare spectra of \cahk\ lines
\citet{MachadanddLinsky1975} showed that the TM in flares is hotter and
formed deeper in the atmosphere than in quiet-Sun models. Similar
results have been found by \citet{Kuridze2017} who constructed SE
models of C-class flare employing high-resolution SST observations of
the \cawav\ and non-LTE inversions.


Spectral diagnostics of thermal conditions near and above TM utilizes
the strong vibration-rotation lines of monoxide carbon (CO) at
4.6\,$\mu$m. 
 Remarkably, the solar CO spectra suggest the coexistence of a
  cool (less than 4000\,K) component of the solar chromosphere with
  the hot, bright gas at 6000 to 7000\,K. Large horizontal velocities
  are observed, implying that the cool component is sustained by the
  supersonic adiabatic expansion of upwelling gas in overshooting
  granules \citep{Solankietal1994}.
This motivated an
introduction of the concepts of thermal bifurcation in the solar outer
atmosphere \citep{Ayres1981} and the so-called COmosphere, a term
coined by G.~Wiedemann \citep{Wiedemannetal1994,Ayres2002}. This zone
probably consists of patchy clouds of cool gas, seen readily in
off-limb emissions of CO lines, threaded by hot gas entrained in
long-lived magnetic filaments as well as transient shock fronts
\citep{Ayres2003,Ayres2010}. The COmosphere was not anticipated in
classical 1D models of the solar outer atmosphere, but is quite at
home in the contemporary 3D highly dynamic view
\citep{Wedemeyeretal2004}.

The 3D R-MHD modeling confirms that the lower solar atmosphere has
much more complex temperature structure where location of TM is not
well-defined \citep[][Figures~12--15]{Bjorgen2019}. However, spatially
averaged temperature structure of these models is still characterized
with a single well-defined TM layer that can be assumed as a boundary
between the photosphere and chromosphere. This raises the questions whether TM is a local, dynamic
plasma parameter rather than a homogeneous global property of the atmosphere. 
Surprisingly, \citet{Wedemeyeretal2004} concluded that TM and the
  upward oriented temperature gradient derived from semi-empirical
  models might be illusive in the sense that they do not really mean
  a rise of the average gas temperature with height.

\begin{figure*}
  \includegraphics[width=0.49\textwidth]{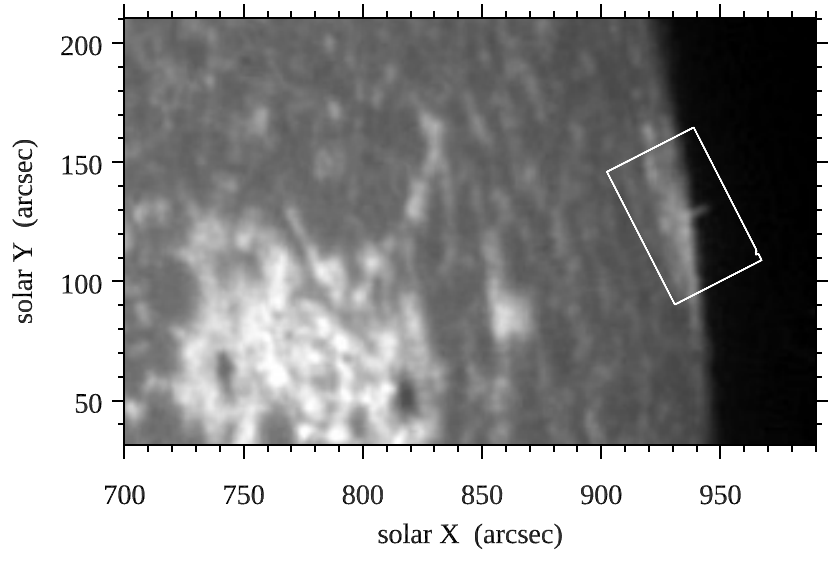}
  \includegraphics[width=0.49\textwidth]{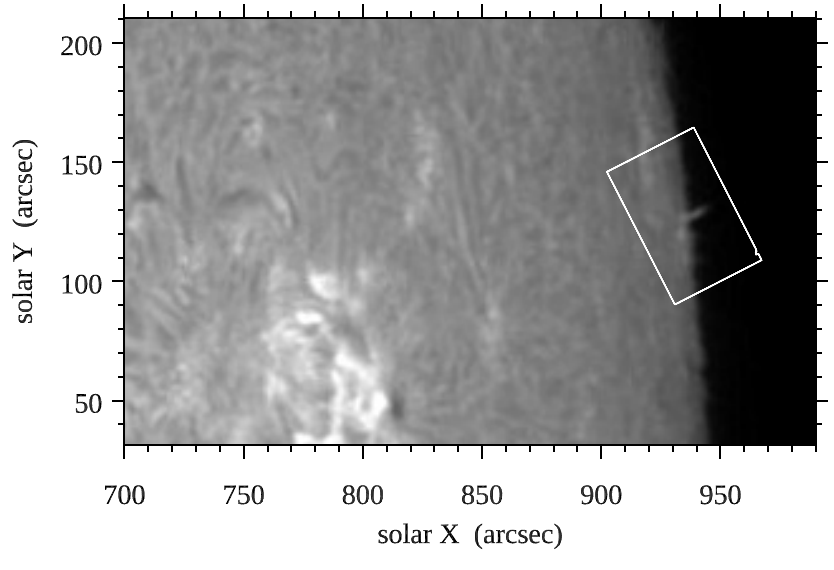}
  \caption{ChroTel \cak\ (left) and \ha\ (right) images, taken on 2018
    June 22 at 08:34\,UT, when the remnants of decaying AR NOAA 12714
    at the limb display the jet at ($x,y$) = (940\arcsec, 130\arcsec),
    which corresponds to the long spicule in the top panels of
    Figure~\ref{fig1} and in the left panel of Figure~\ref{fig4}.}
\label{fig2}
\end{figure*}

One of the key aspects that controls the physical condition in which the emergent intensity forms in the lower atmosphere 
is the radiative coupling between the plasma and underlying radiation. 
The chromosphere is a highly non-LTE environment where radiation is weakly coupled to the local plasma condition. 
Furthermore, the lower atmosphere is highly structured by various processes (convection, jets, magnetic flux tubes, oscillations, waves) 
which appear as large and small scale perturbations of the plasma. Response of the the inhomogeneous, 
dynamic plasma to the underlying radiation defines the complex temperature structure of the atmosphere 
which is manifested in high-resolution, state of the art models and observations.  
Therefore, transitional layer between photosphere and chromosphere can be introduced in terms 
of this interaction rather than simple position of TM. 
Chromosphere can be defined as a region where hydrogen stays predominantly neutral as a result of this complex interaction \citep{Carlsson2007}.

Chromospheric off-limb observations offer a new view on the problem of
TM in association with an off-limb feature called the intensity dark
gap or the intensity dip introduced in \citet{JudgeCarlsson2010}. The
wide-band off-limb filtergrams in the \ha\ center by
\citet{Loughhead1969}, \citet{Nikolsky1970}, and
\citet{AlissandrakisandMacris1971} revealed the dark gap, referred to
as the dark stripe and the dark band in the two latter papers, crowned
with a shell of enhanced chromospheric emission. \citet{Loughhead1969}
suggested that the observed dark gap can be associated with the TM
region of the solar atmosphere \citep{AlissandrakisandMacris1971}.

\citet{JudgeCarlsson2010} analyzed broadband \cah\ filtergrams of limb
spicules obtained by the Hinode's Solar Optical Telescope and pointed
on a dip in the synthetic intensities which is not in qualitative
disagreement with the dip in \ha\ line center intensities discovered
by \citet{Loughhead1969}. Because the relevant formulations by
\citet[][Section~4.2]{JudgeCarlsson2010} are very apt in this context
we quote them literally: ``The cores of \ha\ and neutral helium lines
routinely show a dip in intensity surrounded by a shell of emission
\citep[e.g.,][]{White1963,Loughhead1969,PopeSchoolman1975}. However,
dips seen in visible lines of hydrogen and helium may result more from
the well-known lack of opacity in the low to mid chromosphere and
extra opacity due to fibrils which appear to overarch the stratified
chromosphere. Resolving the issue would require detailed calculations
of \ha\ and He lines with models taking into account the fibril
structure and excitation mechanisms populating these excited atomic
levels.'' However, a few lines further they emphasize:
  ``Importantly, we also note that our calculations never remove the
  off-limb dip entirely \ldots\ yet at least some of the Hinode BFI
  images appear to show no hint of a dip.''

The mentioned helium dip is well visible in the narrow-band filtergram
in the \hed\ line center by \citet[][page~60, 
Figure~4.5]{Libbrecht2016}. The formation of the mid-chromospheric
\ha\ opacity gap is shown in \citet{Leenaartsetal2012}. Recently,
\citet{Paziraetal2017} have detected the dark gap in the \oxy\ line
center with the narrow-band observations obtained with the CRisp
Imaging SpectroPolarimeter (CRISP) on the Swedish Solar Telescope
(SST). In concert with \citet{JudgeCarlsson2010} they concluded that
the dark gap is produced by height variations of the line opacity and
corresponds to the TM layer.

This paper continues in the previous detection and modeling of dark
gap but using high-resolution imaging spectroscopy data of the solar
limb in the \hb\ and \cawav\ lines analyzing them by the
radiative-transfer code RH by \citet{Uitenbroek2001} in the
one-dimension spherical geometry. We study the emergent intensities
from different heights with respect to the photospheric limb and
detect the dark gap between the photosphere and chromosphere in
\hb\ line wing images. We analyze the line contribution functions of
the synthetic spectra and their components to interpret spectral
characteristics of the observed off-limb emissions including formation
of dark gap.

\begin{figure*}
\includegraphics[width=\textwidth]{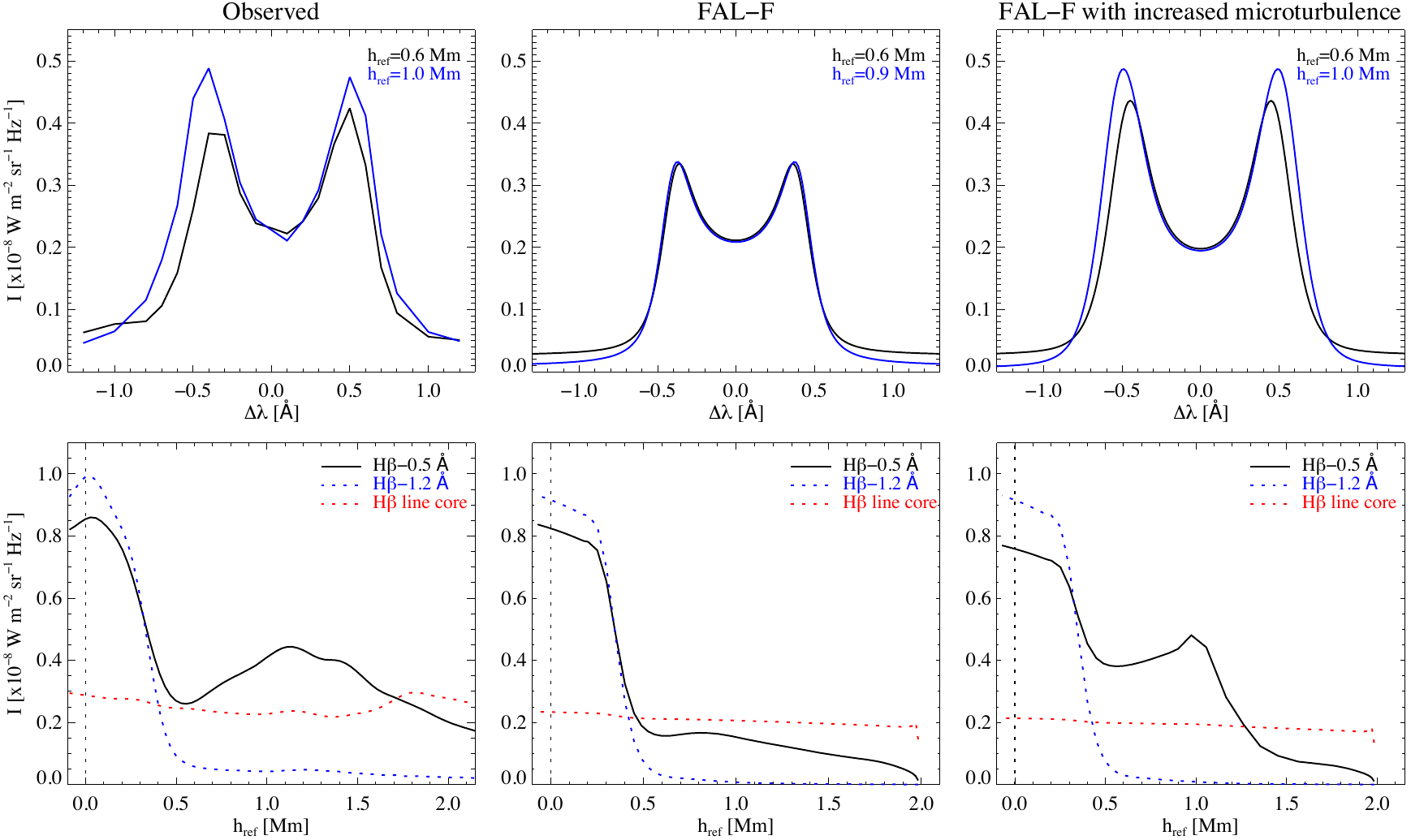}
\caption{Top: Observed (left) and synthetic (middle and right)
    \hb\ line profiles in the dark gap at $h_{\rm ref} \sim 0.6$\,Mm
    (black) and at the maximum of chromospheric emission at $h_{\rm
      ref} \sim 1$\,Mm (blue). Bottom: Observed (left) and synthetic
    (middle and right) monochromatic intensities at indicated
    wavelength separations from the \hb\ line center. The left panels
    pertain to the green dash in the top left panel of
    Figure~\ref{fig1}. The middle and right panels display the
    profiles for standard and increased microturbulence in the FAL-F
    model, respectively.}
\label{fig3}
\end{figure*}

\section{Observations and data reduction}

\subsection{Observational setup}

We observed the western limb of the Sun on 2018 June 22 between 08:20
and 08:43\,UT near the active region NOAA 12714 (Figure~\ref{fig1}).  
The observations were made with the CRISP \citep{Scharmer2006,Scharmeretal2008} and the
CHROMIS \citep{Lofdahleta2021} instruments, both based on dual
Fabry-P\'{e}rot interferometers mounted on the SST
\citep{Scharmeretal2003a,Scharmeretal2003b}.
The CHROMIS observations include narrow-band spectral imaging in the
\hb\ spectral line. The line profile scan consists of 21 profile
samples ranging from $-1.2$\,\AA\ to $+1.2$\,\AA\ at positions $\pm
1.2$, $\pm 1.0$, $\pm 0.8$, $\pm 0.7$, $\pm 0.6$, $\pm 0.5$, $\pm
0.4$, $\pm 0.3$, $\pm 0.2$, $\pm 0.1$, and 0.0\,\AA\ from the line
center. CHROMIS data are processed using the CHROMISRED reduction
pipeline (currently refereed to SSTRED), which includes MOMFBD image
restoration and absolute intensity calibration
\citep{Lofdahleta2021}.
The CRISP data comprises narrow-band imaging spectropolarimetry in the
\cawav\ line profile sampled from $-1.75$\,\AA\ to $+1.75$\,\AA\ in 21
line positions $\pm 1.75$, $\pm 0.945$, $\pm 0.735$, $\pm 0.595$, $\pm
0.455$, $\pm 0.35$, $\pm 0.28$, $\pm 0.21$, $\pm 0.14$, $\pm 0.07$,
and 0.0\,\AA\ from the line center (hereafter, unless specified
otherwise, when referring to the \ca\ line we mean the \cawav\ line).
An acquisition time of each spectral scan of the \ca\ line was
  16\,s but the cadence of the time series was 33\,s due to
  interleaved spectropolarimetric scans in the \fe\ line.
More details on the observations and data can be found in
\citet{Kuridzeetal2021}. In the following we employ data from the
13\,min subinterval of the observational run lasting about 23\,min.

Context imaging of the observed region in the \cak\ and
\ha\ filtergrams (Figure~\ref{fig2}) are provided by the 10 cm robotic
Chromospheric Telescope
\citep[ChroTel,][]{Kentischeretal2008,Bethgeetal2011}, located at the
Observatorio del Teide on Tenerife. ChroTel is a multi-wavelength
imaging telescope for full-disk synoptic observations of the solar
chromosphere in the \cak, \ha, and \heir\ lines.

\begin{figure*}
\includegraphics[width=\textwidth]{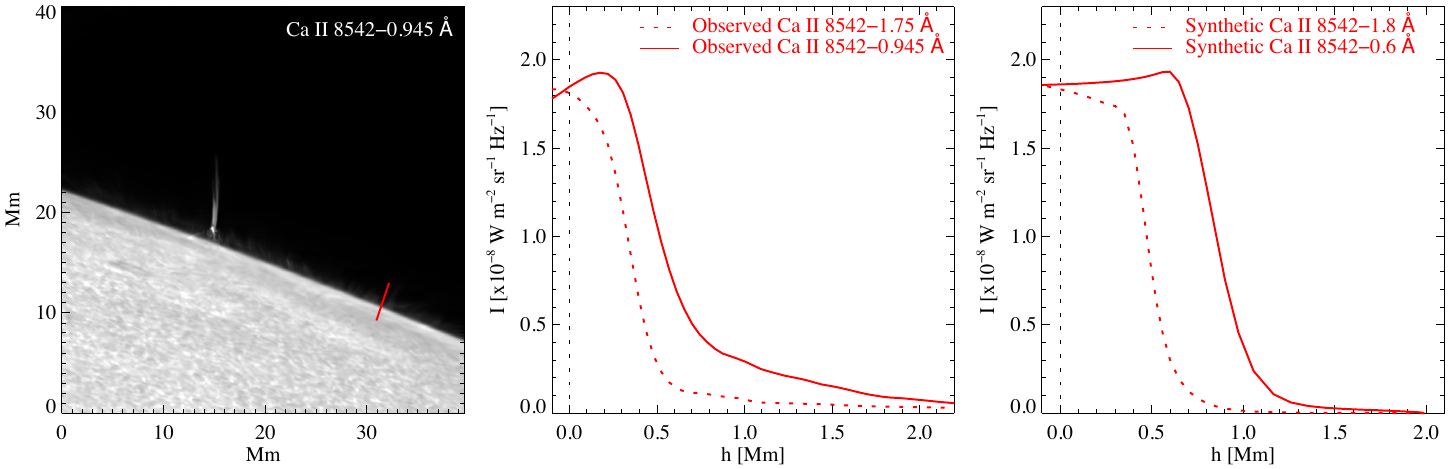}
\caption{CRISP \cawav\ blue wing image at
  $\Delta\lambda=-0.945$\,\AA\ (left). Observed monochromatic
  intensities along the red dash in the left panel (middle) and
  synthetic intensities (right).}
\label{fig4}
\end{figure*}

\subsection{Radiometric calibration and convolution} 

The comparison of SST data with outputs of radiative transfer
computations requires careful intensity calibration of the former and
convolution of the latter with the transmission profiles of the SST
instruments provided by J.~de la Cruz Rodr\'{\i}guez (2017, private
communication) and M.~L\"{o}fdahl (2019, private communication). Here
we use the same CRISP data as in the paper by \citet{Kuridzeetal2021}
and its Section~2.2 ``Data Radiometric Calibration'' is also valid for
this work. The processing pipeline yields the CHROMIS data implicitly
calibrated in disk-center absolute intensity units using the Hamburg
disk center spectral atlas as a reference \citep{Neckel1999},
disregarding the actual position angle of CHROMIS science data
\citep[][Section~4.4]{Lofdahleta2021}. Therefore we perform
re-calibration of the CHROMIS data following the same procedure as at
CRISP data but taking the \hb\ profiles from \citet{David1961} as a
calibration reference and the CHROMIS transmission profile provided by
M.~L\"{o}fdahl (2019, private communication). The resulting
intensities expressed relative to the quiet-Sun continuum intensity at
the disk center are converted to the absolute intensities in the
radiometric units \WHz\  by the disk-center absolute continuum
intensity taken from \citet{Cox2000}.

\section{Analyses and results}

\subsection{Appearance of \hb\ dark gap}
\label{appearance}

The top panels of Figure~\ref{fig1} show the western limb in the
\hb\ line wing images at $\Delta\lambda=\pm0.5$~\AA\ taken on 2018
June 22 at 08:34:24\,UT. The SE model of the spicule in
the center of the Field of View (FoV) is presented in
\citet{Kuridzeetal2021}.  This work is aimed at the off-limb emissions
from the edge of photosphere up to the chromosphere.  Both \hb\ wing
images suggest a presence of dark gap which closely follows the limb
below the chromospheric canopy and the forest of type II spicules. To
investigate the location of dark gap the intensities at
different time moments are extracted along the three radial dashes of
length 6\,Mm shown in the top left panel of Figure~\ref{fig1}. The
length of these dashes is chosen to cross the limb, the dark gap, and
the off-limb chromosphere at different parts of FoV.  The
time-distance diagrams along each dash (middle panels of
Figure~\ref{fig1}) and density diagrams produced by superposition of
extracted intensities at different time moments (bottom panels of
Figure~\ref{fig1}) clearly confirm the existence of dark gap at the
selected portions of FoV over most of the 13\,min time span.

We define the observable photospheric limb in the \hb\ imagery
(hereafter as the apparent limb) as time-averaged positions of
inflection points of the steepest intensity gradients along the dashes
in the \hb\ far wing image at $\Delta\lambda =-1.2$\,\AA\ (not
shown). The position of the apparent limb defines the zero point of
the spatial scales at $x$ axes in the bottom panels of
Figure~\ref{fig1}. We will refer to the spatial scale as the apparent
height denoted as $h_{\rm app}$. However, the position of the
observable photospheric limb slightly varies possibly due to
small-scale photospheric inhomogeneities producing local Wilson
depressions. Therefore the apparent height of dark gap also slightly
varies along the observable photospheric limb. It is shown in the top
right panel of Figure~\ref{fig1} in which the red contour follows the
positions of dark gap intensity minima yielding the average apparent height of
$0.265 \pm 0.055$\,Mm. We take this value as the height of dark gap
above the apparent limb.

We note that the apparent limb is uplifted by about $~0.35$\,Mm above
the level with the optical depth unity at 5000\,\AA\ viewed at the
disk center. More details on the subject can be found in
\citet[][pages 8--9, 279--280]{Athay1976} and
\citet[][Table~1]{Lites1983}. This off-set is due to an increase of optical depth at 5000\,\AA\ when
moving LoS from disk center ($\mu=1$) to limb ($\mu=0$).
  
Example of the observed monochromatic intensities along the green dash
are shown in the bottom left panel of Figure~\ref{fig3}. Here we
introduce the reference height $h_{\rm ref}$ shifted with respect to
$h_{\rm app}$ as $h_{\rm ref} = h_{\rm app} + 0.35$\,Mm. It will be
useful for later comparisons with the synthetic intensities computed
by the RH code which will refer to the base of the photosphere
($h_{\rm ref}=0$ in the bottom middle and right panel of
Figure~\ref{fig3}) with the optical depth unity at 5000\,\AA. Now the
bottom left panel of Figure~\ref{fig3} shows the dark gap at
$\Delta\lambda =-0.5$\,\AA\ at $h_{\rm ref} \sim 0.6$\,Mm compared to
$h_{\rm app} \sim 0.265$\,Mm in the bottom panels of
Figure~\ref{fig1}. Remarkably, the \hb\ line core intensity lacks any
signature of dark gap compared to the previous \ha, \hed, and
\oxy\ detection referred to in Section~\ref{intro}. The top left panel
of Figure~\ref{fig3} shows the \hb\ line profiles at the dark gap at
0.6\,Mm and at the point of Maximum Off-Limb Emission (hereafter as
MOLE in the context of observed and synthetic intensities) outside the
dark gap where the observed off-limb emission has its local maximum at
the height 1\,Mm. For example, the contrast of dark gap is apparent by
comparing the profile intensities at, e.g., $\Delta\lambda = \pm
0.5$\,\AA. Generally, the \hb\ wing intensities of the dark gap
profile at 0.6\,Mm are reduced over the wavelength spans
$(-0.9,-0.3)$\,\AA\ and $(0.4, 1.1)$\,\AA\ compared to the MOLE
profile.  \hb\ reversed core in the dark gap is very close to the MOLE
profile.

The left panel of Figure~\ref{fig4} shows the \ca\ blue wing image at
the wavelength position $\Delta\lambda=-0.945$\,\AA\ from the line
center taken on 2018 June 22 at 08:34:27\,UT at the western limb. The
middle panel shows the monochromatic intensities extracted along the
red dash in the left panel. The image and the intensity extractions at
all wavelength positions across \ca\ line scan lack any trace of dark
gap. Here the observable photospheric limb is also defined by the
position of inflection point of intensity gradient along the red dash
in the \ca\ far wing image at $\Delta\lambda =-1.75$\,\AA\ (not
shown). The position shifted about 0.35\,Mm defines the
height $h$ used at $x$ axes in the middle and right panels. Here we
omit the adjective 'reference' and the subscript $_{\rm ref}$ to
distinguish $h$ from its \hb\ and RH counterpart $h_{\rm ref}$.
This is because the end wavelength settings of the \ca\ scans at
$\Delta\lambda=-1.75$\,\AA\ do not reach yet a true photospheric
continuum thus introducing an unknown but likely small off-set
between the $h$ and $h_{\rm ref}$ scales. Remarkably, the middle panel
of Figure~\ref{fig4} clearly shows a small intensity enhancement at
$\Delta\lambda=-0.945$\,\AA\ over the height span from 0.1 to
0.25\,Mm. Such enhanced off-limb intensities are observed within the
wavelength spans $(-0.945,-0.6)$\,\AA\ and $(0.6,0.945)$\,\AA\ of
\ca\ profile.

\begin{figure}
\begin{center}
\includegraphics[width=0.485\textwidth]{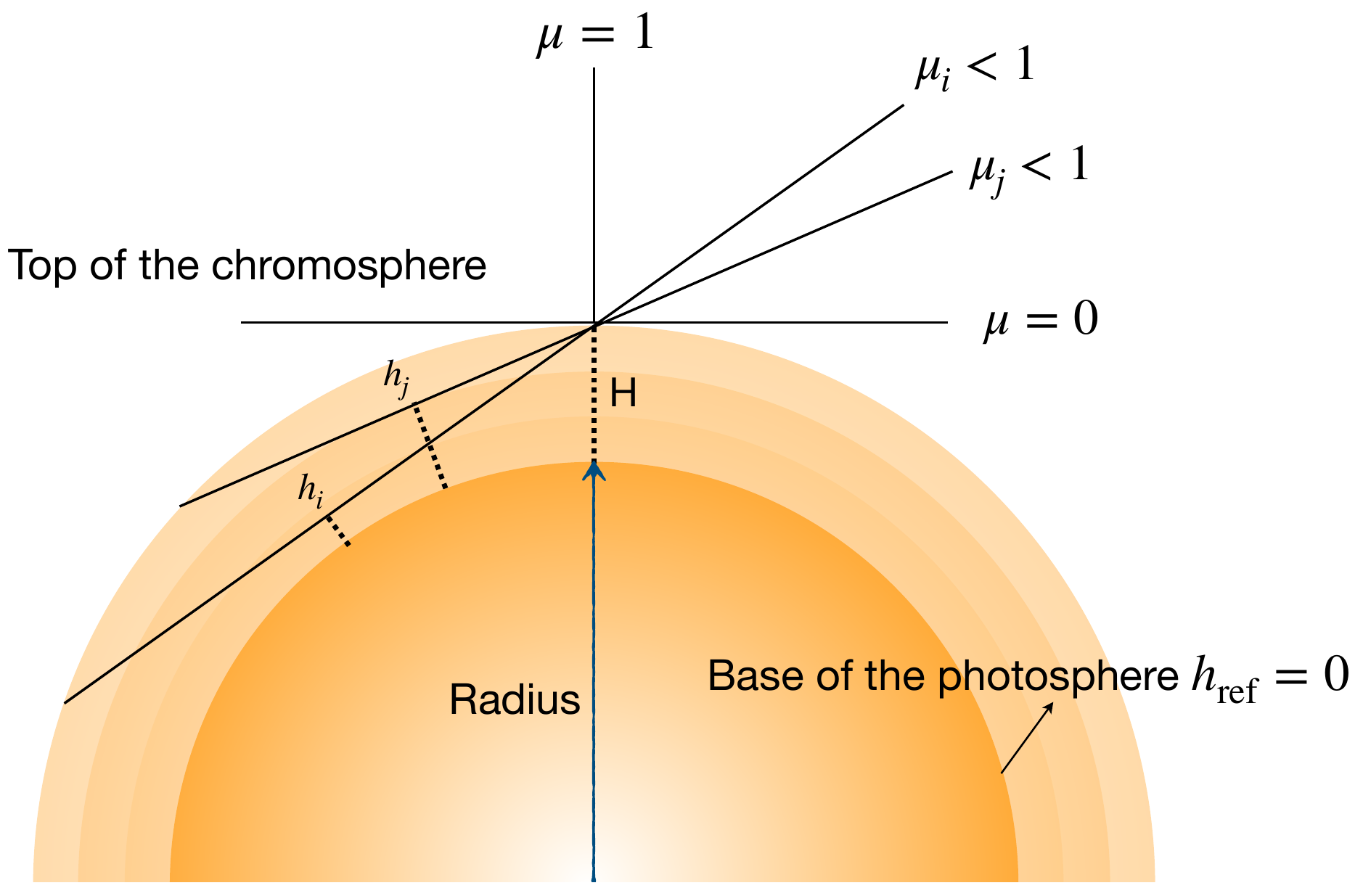}
\caption{Sketch of off-limb geometry showing the line-of-sight rays
  passing at different heights $h_i$ and $h_j$ above the base of the photosphere. 
  The RH code computes emergent intensities at the top of
    the model atmosphere. As a result {$\mu = 0$ corresponds to
      the emergent intensity at the height $H = 2$\,Mm above the base of
      the photosphere at $h_{\rm ref} = 0$}.}
\label{fig5}
\end{center}
\end{figure}


\section{RH code and viewing geometry}

\begin{figure*}
\includegraphics[width=\textwidth]{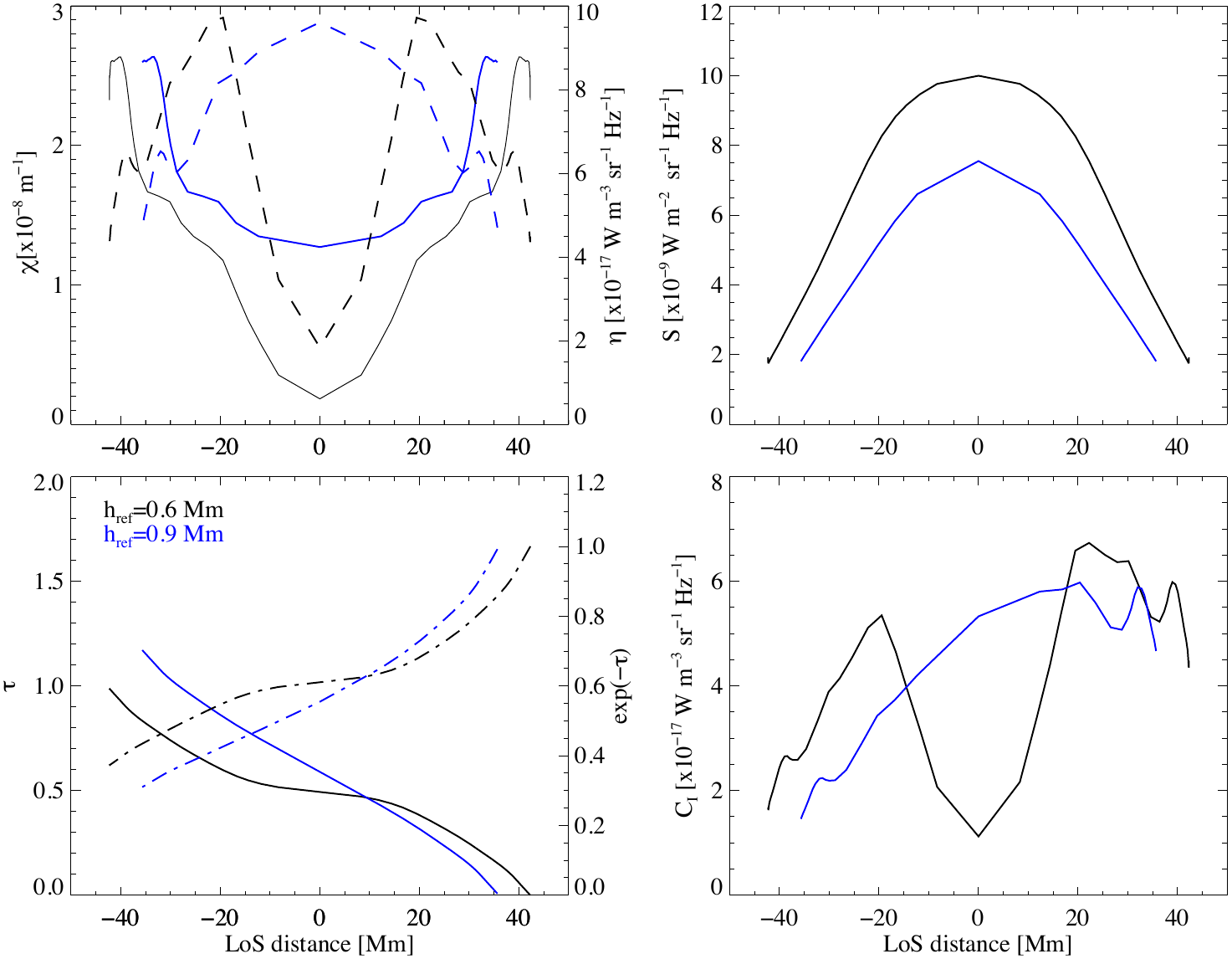}
\caption{Formation characteristics of the \hb\ blue wing intensities
  at $\Delta\lambda = -0.44$\,\AA\ shown along the line of sight in the
  dark gap at the reference height $h_{\rm ref} \sim 0.6$\,Mm (black)
  and at the maximum off-limb emissivity at 0.9\,Mm (blue). Top left:
  Opacity $\chi$ (left $y$ axis) and emissivity $\eta$ (right $y$
  axis) as the solid and dashed lines, respectively. Top right: The
  source function $S$. Bottom left: The optical depth $\tau$ (left $y$
  axis) and the exponential $\exp(-\tau)$ (right $y$ axis) shown as
  the solid and dash-dotted lines, respectively. Bottom right: The
  intensity contribution function $C_I$.}
\label{fig6}
\end{figure*}

To interpret the observational characteristics of the off-limb
\hb\ and \ca\ line emissions we generated synthetic line profiles with
the radiative transfer code RH by \citet{Uitenbroek2001}.  RH can
calculate spectral line profiles for a model atmosphere by solving the
equations of statistical equilibrium and radiative transfer under the
non-LTE conditions.  We utilized the one-dimension spherical version
of RH code.  Spherical geometry takes into account the tangential
variation of physical parameters along the line of sight (LoS),
including optical depth which is the main quantity for radiative
transfer calculations. This makes spherical RH the most appropriate
choice for calculations of off-limb emissions.

To synthesize the line profiles, we chose the SE model of a bright
area of the network FAL-F \citep{Fontenlaetal1993} extending about
2\,Mm from the base of the photosphere to the top of the
chromosphere. This choice was motivated by an assumption that the
formation of the \hb\ and \ca\ lines in the dark gap is affected by
the remnants of decaying AR NOAA 12714 seen at the limb in the context
\cak\ and \ha\ images (Figure~\ref{fig2}) taken by the ChroTel.

 We used the models of the hydrogen and singly ionized calcium
  atom with the five bound levels plus continuum with the complete
  frequency redistribution for the former (apart from the \lya\ and
  \lyb\ transitions which were calculated in partial frequency
  redistribution (PRD)) and the PRD with the cross-redistribution for
  Raman scattering for the latter.

The RH computes emergent intensities at the top of the model
atmosphere at 81 pre-defined direction cosines $\mu_i$, indexed as
$i=1\dots81$, from the disk center {at $\mu_{81} = 1$
to the top of the chromosphere} at $\mu_1
\sim 0$ assumed at the height $H = 2$\,Mm above the base of the
  photosphere. Figure~\ref{fig5} shows a sketch of viewing geometry
of different LoS rays passing closer to $(\mu_i, h_i)$ or
further from $(\mu_j, h_j)$ the base of the photosphere 
corresponding
  to the reference height $h_{\rm ref}=0$. Note that $\mu_i>\mu_j$
and $h_i<h_j$ in this sketch.  Therefore, in our RH calculations $\mu_1 \sim 0$
  and $\mu_{67}\sim0.078$ correspond to the tangential rays touching
  the top of the chromosphere and the base of the photosphere,
  respectively. Thus the rays along $\mu_i$ with $0<i<67$ represent
emissions emerging from different heights between $h_{\rm ref}\sim 0$
to $\sim 2$\,Mm (Figure~\ref{fig5}) in the
direction of LoS. The conversion formula between the pre-defined
  direction cosines $\mu_i$ and the corresponding reference heights
  $h_{\rm ref,i}$ reads
\begin{equation}
h_{\rm ref,i} = (\nom{R} + H)\sqrt{1 - \mu_i^2} - \nom{R}\,,
\label{eq1}  
\end{equation}
where \nom{R} is the nominal solar radius.

\begin{figure*}
\includegraphics[width=\textwidth]{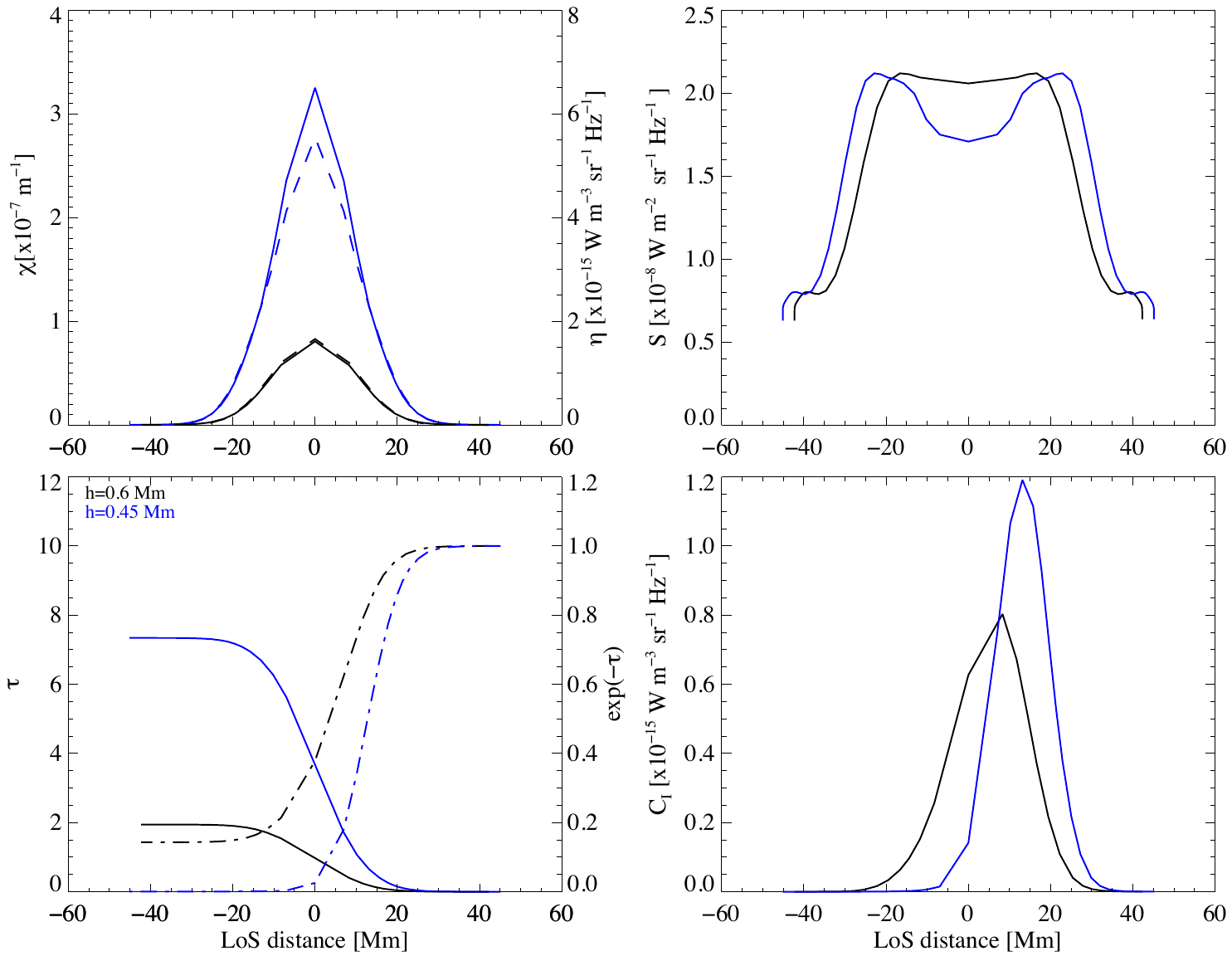}
\caption{Formation characteristics of the \cawav\ blue wing
  intensities at $\Delta\lambda = -0.6$\,\AA\ shown along the line
  of sight at the heights $h \sim 0.45$\,Mm (blue) and 0.6\,Mm
  (black). Panel descriptions are the same as in Figure~\ref{fig6}.}
\label{fig7}
\end{figure*}

\subsection{Synthetic \hb\ and \cawav\ line profiles}

The synthetic emergent monochromatic intensities of the \hb\ line
along off-limb LoS rays as a function of the reference height $h_{\rm
  ref}$ (Section~\ref{appearance}) are presented in the bottom middle
and right panels of Figure~\ref{fig3} for the three spectral
positions. Middle panels show intensities for FAL-F atmosphere with
standard microturbulence velocity input.  While intensity dip (i.e.,
the dark gap) is detected at the height of 0.6\,Mm at the blue wing
spectral position of $\Delta\lambda=-0.5$\,\AA\ there are no intensity
variations in the line core (bottom middle panel of
Figure~\ref{fig3}). The top middle panel of Figure~\ref{fig3} shows
the synthetic \hb\ dark gap and MOLE profiles at $h_{\rm ref} \sim
0.6$\,Mm and 0.9\,Mm, respectively. It demonstrates that the wing
intensities of the MOLE profile are only slightly enhanced over the
wavelength spans $|\Delta\lambda|\in(0.35,0.55)$\,\AA\ compared to the
\hb\ line wing intensities at the height of 0.6\,Mm. It produces a
synthetic contrast observable as the dark gap.

We have also synthesized line profiles for FAL-F model with increased
microturbulence in the chromosphere (right panels of
Figure~\ref{fig3}). In FAL-F, at height range covering $\sim 0.8 -
1.4$\,Mm $\vmic$ varies between $\sim 1$ and 5\,\kms. We took
$\vmic$($h_{\rm ref}=0.8 - 1.5$\,Mm)$\sim 4 - 13$\,\kms\ in the FAL-F
model. Here we use quite different viewing angle (off-limb) than in SE
models like FAL (on-disk) and we assume that the projected LoS
velocities have a larger scatter.  Resulted line profile and intensity
as a function of height is presented in the right panels of
Figure~\ref{fig3}. The increased microturbulence produces broader line
profiles and higher contrast between the dark gap and MOLE line wing
intensities. Furthermore, gap is formed over a wider wavelength range
extending toward the outer wings at $|\Delta\lambda|\in(0.4,
0.85)$\,\AA\ (top right panel of Figure~\ref{fig3}).

In contrast to \hb, the synthetic monochromatic emergent intensities
in the \ca\ line wings (right panel of Figure~\ref{fig4}) along
off-limb LoS rays do not show the dark gap but the intensity
enhancements at $h \sim 0.6$\,Mm within the wavelength spans
of $(-0.3,-0.6)$\,\AA\ and $(0.3, 0.6)$\,\AA.
  We note that the
  results presented in middle and right panels of Figure~\ref{fig3}
  and also in the right panel of Figure~\ref{fig4} are convolved with
  the instrumental characteristics of the CHROMIS and CRISP
  instruments, respectively.

Finally, the results (synthetic profiles, appearance of the dark
gap) obtained through RH calculations with the FAL-F model atmosphere
are very similar to those by the FAL-C model (an average region of the
quiet Sun).  However, due to the better match between absolute
observed and synthetic intensities (compare the left and right panels
in Figure~\ref{fig3}) we give priority to the FAL-F model in this
work.  Our selection is also justified with context images of the
larger FoV confirming that observed limb contains chromospheric bright
network, which makes FAL-F more appropriate model for the observed
data (Figure~\ref{fig2}.)

\subsection{Understanding the dark gap formation}
\label{cf}

To understand the off-limb \hb\ and \ca\ emissions, we examine their
intensity contribution functions $C_I$ at selected gap-pertinent
wavelengths as a function of distance along off-limb LoS rays.
Emergent intensity $I_\nu$ can be expressed in terms of contribution
function \citep{CarlssonandStein1994},
\begin{equation}
I_\nu = \int_{\rm LoS}C_I\,dl = \int_{\rm LoS}S_\nu\,\chi_\nu\exp(-\tau_\nu)\,dl\,,
\label{eq1}
\end{equation}
where $S_\nu$ is the source function at the frequency $\nu$,
$\chi_\nu$ is the opacity, 
$\tau_\nu$ is the optical depth, and $l$ is the
geometrical distance along LoS. Its zero reference point corresponds
to the cross point of viewing LoS ray with the normal to both the
photospheric surface and the ray (Figure~\ref{fig5}). The positive $l$
values increase toward the observer.

Figure~\ref{fig6} shows the intensity contribution function $C_I$ of
the FAL-F atmosphere with standard microturbulence input for the
\hb\ blue wing at $\Delta\lambda = -0.44$\,\AA\ (bottom right panel)
and its components at the dark gap (black curves) and at the point of
MOLE (blue curves). To facilitate reading, the panel references omit
Figure~\ref{fig6} in this paragraph. For the ray passing through the
dark gap the line wings have the lowest opacity $\chi$ and emissivity
$\eta$ near the limb at $l=0$\,Mm (black solid and black dashed
curves, respectively, in top left panel). The optical depth unity is
reached at the farthest point of the dark gap atmosphere at $l \sim
-40$\,Mm (black curve in bottom left panel). However, due to the low
emissivity $\eta$ of the dark gap atmosphere (black dashed in top left
panel) there is a very little contribution to the emergent intensity
over the LoS span $(-10,10)$\,Mm (black curve in bottom right
panel). The ray passing through the point of MOLE at 0.9\,Mm is
characterized with a slightly higher optical depth $\tau$ of about 1.2
(blue curve in bottom left panel). However, a high emissivity $\eta$
at this point over the LoS span $(-10,10)$\,Mm (blue dashed in top
left panel) yields a higher emergent intensity than in the dark gap
(blue curve in bottom right panel).

Figure~\ref{fig7} shows the intensity contribution function $C_I$ for
the \ca\ blue wing at $\Delta\lambda = -0.6$\,\AA\ (bottom right
panel) and its components at the point of MOLE at 0.6\,Mm (black
curves, see also right panel of Figure~\ref{fig4}), coinciding with
the height of \hb\ dark gap, and at the height of 0.45\,Mm (blue
curves) selected for comparison. $C_I$ components indicate that ray
passing through the atmosphere at $\sim 0.6$\,Mm experiences low
opacity $\chi$, low emissivity $\eta$, and optical depth $\tau$ (top
left and bottom left panels of Figure~\ref{fig7}) resulting in
optically thin emission. 
At $\sim 0.45$\,Mm the rays
originate in an optically thicker environment at $\tau \sim 7$
attenuating the emissivity through the factor $\exp(-\tau)$. As a
result, the off-limb emission at \ca\ line wing, emerging at $\sim
0.6$\,Mm, has higher intensity than at $\sim 0.45$\,Mm.

\begin{table}
  \centering
  \caption{Parameters of prominent chromospheric lines.}
  \label{tab1}
  \begin{tabular}{l r D D}
    \hline
    \hline
    Atom/Ion & $\lambda$\,(\AA)  &  \multicolumn2c{$E_{\rm low}$\,(eV)} & \multicolumn2c{$E_{\rm low}/k_{\rm B}T_{\rm e}$\tablenotemark{\footnotesize a}} \\
    \decimals
    \hline
    \hed         &  5876  &  21.0  & 34.8  \\
    \he          & 10830  &  19.8  & 32.8  \\
    \ha          &  6563  &  10.2  & 16.9  \\
    \hb          &  4861  &  10.2  & 16.9  \\
    \oo          &  7772  &   9.1  & 15.1  \\
    \ca          &  8542  &   1.7  & 2.8  \\
    \cak         &  3934  &   0.0  & 0.0  \\
    \cah         &  3968  &   0.0  & 0.0  \\
    \hline
  \end{tabular}
  \tablenotetext{a}{computed for a typical chromospheric electron temperature $T_{\rm e} \sim 7000$\,K.}
\end{table}

\subsection{Understanding dark gap visibility didactically}

In the previous Section~\ref{cf} we attempted to render a rigorous
explanation of dark gap visibility in the context of the TM
region. Here we offer a more straightforward account for why the dark
gap is visible in \hb\ and other chromospheric lines, listed in
Section~\ref{intro}, but not in \ca.

The line opacity is proportional to the fraction of atoms (or ions)
excited to the $n$th level which scales with the Boltzmann factor as
${\rm const} \times g_{\rm low}\,{\rm e}^{-E_{\rm low}/k_{\rm B}T_{\rm e}}$,
where $g_{\rm low}$ and $E_{\rm low}$ is the statistical weight and
energy of lower level of transition, respectively, $k_{\rm B}$ is the
Boltzmann constant, and $T_{\rm e}$ is the electron temperature
\citep{Gray2008}. Table~\ref{tab1} shows the lower level energies and
the exponents of the Boltzmann factor for $T_{\rm e} \sim 7000$\,K for
prominent chromospheric spectral lines. The lines with $E_{\rm low}
\gg 1$\,eV have high Boltzmann factor exponent suggesting they should
have low opacities in the chromosphere. This explains why the dark gap
is well visible in the \he, Balmer, and \oo\ lines but hardly or not
 at all in the \ca\ line.

However, the \cahk\ lines throw uncertainty on the proposed scenario
which suggests that a presence or absence of dark gap is controlled by
the $E_{\rm low}$ of transition. The model calculations of \cah\ by
\citet[][Figure~5]{JudgeCarlsson2010} never remove the off-limb dip
entirely which should not appear at all like in our observations and
calculations of \cawav.

\section{Discussion}

The wing detection of the dark gap and its absence in the \hb\ core is
particularly significant in the context of previous gap detection in
\ha, \hed, and \oxy\ reviewed in Section~\ref{intro}. The gap
detection in \ha\ pertained to wide-band observations blending the
line center and wing intensities. High inherent opaqueness of the
chromosphere in the \ha\ center enhanced further by the spicules and
the chromospheric canopy, prevents observing the dark gap in the
\ha\ center. We consider the past \ha\ detection in fact for wing
detection in the regime of low \ha\ optical thickness outside the
thick opacity wall due to the spicules and the canopy. On the other
hand, the recent narrow-band gap detection by the SST/CRISP in the
\hed\ and \oxy\ centers give evidences about low inherent optical
thickness of the chromosphere \citep[][$\tau$ in 
Figure~7]{Paziraetal2017}, the spicules and the canopy in these
lines.  We note that in \citet{Paziraetal2017} there is around 170\,km
difference between the location of the dark gap in observed and
synthetic \oxy\ line core intensities
\citep[][Figure~6]{Paziraetal2017}.  For \hb\, analyzed in this work,
both the observed and synthetic gaps, computed by the FAL-F and FAL-C
models, are located at $\sim 0.6$\,Mm suggesting that \hb\ line wing
images could be better tracer of TM layer.

There are noticeable differences between properties of the dark gap in
observations and the RH radiative transfer simulations. In particular,
(\RN{1}) the observed contrast of the dark gap with respect to maximum
off-limb emission is larger in observations than in the RH simulations
(bottom panels in Figure~\ref{fig3}); (\RN{2}) the observed
\hb\ profiles are broader and asymmetric compared to the synthetic
profiles (top panels in Figure~\ref{fig3}); (\RN{3}) dark gap is seen
along a wider spectral range in the observation covering around
$|\Delta\lambda|\in(0.4 - 1)$\,\AA, at the blue and red wings, whereas
with standard FAL-F microturbulence velocities, gap is only detected
at $|\Delta\lambda|\in(0.35,0.55)$\,\AA.  We explain these differences
as due to absence of a realistic model of spicules and the
chromospheric canopy in the RH simulations. Spicule emission
contributes to the observed emergent intensities in the layers above
the dark gap thus enhancing the observed contrast. Spicule hydrogen
profiles of the Balmer series are broad with enhanced and asymmetric
wings \citep{Beckers1972,Shojietal2010}. It likely makes the observed
\hb\ profiles broad and asymmetric as captured in our observations. To
investigate the effect of extra broadening introduced by chromospheric
dynamics such as spicules, flows, jets, MHD waves, we performed a
simple test by using increased microturbulence in FAL-F model at the
heights of $0.8 - 1.5$\,Mm above the photosphere.  Resulted profiles
showing broad line width and presence of the dark gap at the outer
wing positions at $|\Delta\lambda|\in(0.4,0.85)$\,\AA, are in a good
agreement with the observations (top left and right panels of
Figure~\ref{fig3}).

The dark gap is absent in \ca, both in observations and simulations
(Figure~\ref{fig4}). As discussed in \citet{Cauzzietal2008}, an
opacity gap in the \ca\ line wings is less prominent due to the much
lower excitation energy of 1.7\,eV of the D term of the \ca\ infrared
triplet compared to the excitation energy of hydrogen upper levels
10.2\,eV (Table~\ref{tab1}). 
Furthermore, the low optical depth of the \ca\, line wing emission produces higher off-limb synthetic emergent intensity at $\sim
0.6$\,Mm than at $\sim 0.45$\,Mm from the photospheric limb (Figure~\ref{fig4} and \ref{fig7}).

\begin{figure}
\centering
\includegraphics[width=0.47\textwidth]{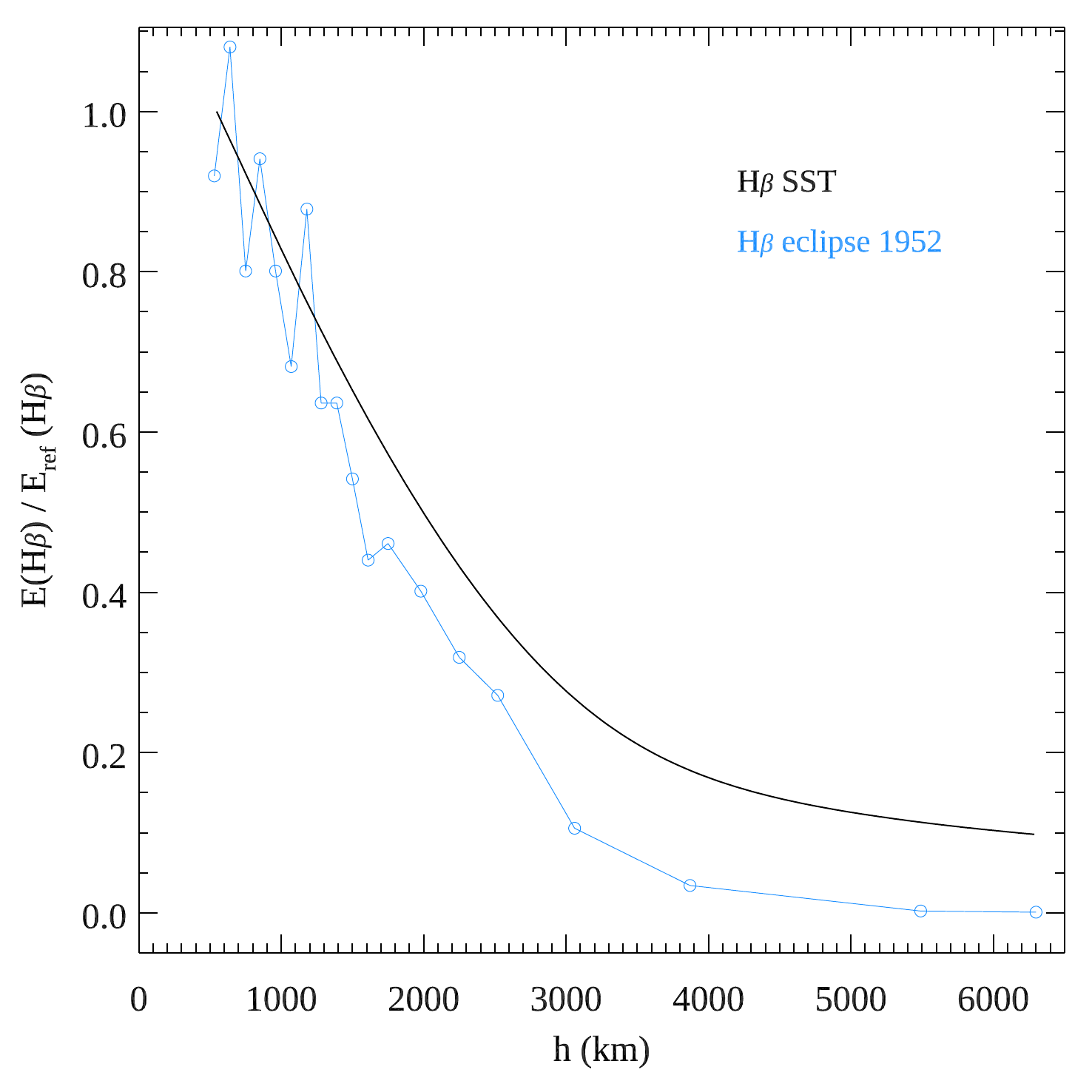}
\caption{
  Relative \hb\ height gradients, $E/E_{\rm ref}$. The
  wavelength- and area-integrated intensity $E$ of the \hb\ line as a
  function of the height above the photospheric limb $h$ derived from
  the SST/CHROMIS spectral imaging at the limb on 2018 June 22 (black)
  and from the flash spectra (blue) obtained at the 1952 eclipse in
  Khartoum \citep[][Table~2]{Athayetal1954}. The SST intensities $E$
  are normalized with respect to ad-hoc reference intensity $E_{\rm
    ref}$ at $h = 530$\,km. The eclipse intensities $E$ are normalized
  with respect to the ad-hoc reference value $E_{\rm ref}$ taken as an
  average of the intensities at the heights 530\,km and 640\,km.
  }

\label{fig8}
\end{figure}

In the past, eclipse flash spectra provided a sound basis for
determining physical conditions in the chromosphere
\citep{Athayetal1954,Judgeetal2019}. Out of the wealth of continuum
and line spectral data presented therein we concentrate further just
on the \hb\ data \citep[][Table~2]{Athayetal1954}. The \hb\ flash
spectra sampled the quiet-Sun chromosphere over the heights from
530\,km up to 6300\,km above the photospheric limb yielding twenty
samples with the regular cadence of 0.4\,s and step size of 110\,km in
the lower chromosphere but increasing higher up. Notice the blue
circles in Figure~\ref{fig8} comparing the wavelength- and
area-integrated \hb\ intensity $E$ as a function of the height $h$ above
the photospheric limb derived from the SST/CHROMIS data (black) and
the 1952 eclipse flash spectra shown also in \citet[][page~177,
  Fig.~V-7.]{Athay1976}. Both gradient curves compare well in the low
chromosphere up to $h \sim 2000$\,km. Higher up they bear on spicules
as broadly discussed in \citet{Athay1976}. The SST curve indicates
relatively brighter spicules than those captured by the eclipse flash
spectrum. This is in full agreement with the results by
\citet[][Figure~5]{JudgeCarlsson2010} showing brighter spicules in the
\cah\ images acquired in the polar coronal hole than those captured in
the flash spectrum. It was shown in \citet[][Chapter~II]{Athay1976}
that when even 0.1 to 1\% of the disk area is covered with
spicule-like features the effects at the limb are dramatic. It is
entirely possible, and probable as well, that nearly all of the
emission at the limb above heights of 1500\,km arises in spicules
covering less than 1\% of the solar surface
\citep[][page~305]{Athay1976}. These same spicules would have quite
negligible effect on much of the emission observed near the disk
center \citep{Judgeetal2020}. Assuming that the limb is completely
covered by spicules at $h \sim 3000$\,km and that at $h \sim 5000$\,km
the mean fraction of the limb covered by spicules is approximately
10\%, \citet[][pages~178--179]{Athay1976} suggested that most of the
decrease in \hb\ brightness above 3000\,km is due to the decreasing of
surface area of spicules. The plausibility of this suggestion is
demonstrated in Fig.~VII--9 by \citet[][page~305]{Athay1976} showing
the relative decrease of spicule numbers and the \hb\ emission. Thus,
the spicule number distribution falls off more slowly with height than
does \hb\ emission \citep[][page~305]{Athay1976}. It follows that if
the spicules are effectively thin and have outwardly decreasing
material density they can account for the observed \hb\ line emission
above a height of about 2500\,km \citep[][pages~305--306,
  Fig.~VII--10]{Athay1976}.

\citet{Judgeetal2020} examined possible candidates of the differences
in central reversal (canter-to-peak ratio) between observed and
synthetic chromospheric lines such as \lya, \mghk, \cahk.
They analyzed disk profiles and concluded that micro and macro
turbulence are not responsible for the observed discrepancies.  They
also admitted that the area coverage of spicules is too small to
contribute significantly to the spatially averaged profiles (page~4
therein, left column).  However, as it has been mentioned above the
effects of spicules at the limb are dramatic
\citep[][page~305]{Athay1976} as more spicules are expected to be
aligned along the LoS. Our motivation to invoke enhanced
microturbulence was sparked by the results presented in
\citet{Teietal2020} proving a profound effect of microturbulence on
the \mghk\ profiles of spicules computed by multi-slab non-LTE
modeling. A question may arise whether macroturbulence may account for
the dark gap as well. An effect of macroturbulence on a line profile
is formally represented by a convolution affecting inevitably the
whole line profile including its core. However, as we show in
Figures~\ref{fig1} and \ref{fig3} the dark gap pertains only line
wings and not the core. This excludes macroturbulence from any
considerations and keeps microturbulence as a viable representation of
multitude of spicules at the limb in radiative transfer calculations
by the spherical RH code.

\section{Conclusions}

We have analyzed high-resolution off-limb imaging spectroscopy in the
\hb\ and \cawav\ lines and compared them with radiative transfer
computations. The \hb\ line wing images and intensity profiles of
off-limb wing emissions show the dark gap at the height about
0.265\,Mm above the apparent photospheric limb (Figure~\ref{fig1})
corresponding to the height 0.6\,Mm above the base of the photosphere
(bottom left panel of Figure~\ref{fig3}).  The gap disappears in the
line core and far wing images at $|\Delta\lambda|\gtrsim 1$\,\AA.  Due
to a spherical symmetry of the solar atmosphere, off-limb LOS
chromospheric rays are passing though a very long ($\sim 100$\,Mm)
geometrical path length covering very dynamic and non-homogeneous
chromospheric layers.  Therefore, to understand the formation of the
dark gap radiative transfer calculations need to be performed for
off-limb emissions in spherical geometry.

The synthetic \hb\ profiles, computed by the RH code and the FAL-F
model along off-limb line-of-sight rays, confirm the existence of the
dark gap at the height of 0.6\,Mm, which is close to the temperature
minimum region of the FAL-F atmosphere at about 0.525\,Mm. The
temperature of this region $\sim 4500$\,K is too low to populate upper
hydrogen levels resulting in the opacity and emissivity gap
\citep{Cauzzietal2008,Leenaartsetal2012}.  The off-limb line-of-sight
ray, passing at the height of $\sim 0.6$\,Mm, samples a large segment
of the TM region with meager population of relevant \hb\ levels
(Figure~\ref{fig5}).  The components of the intensity contribution
function show that a strong opacity and emissivity gap in this layer
makes the local plasma and the \hb\ line wing emission at e.g.,
$|\Delta\lambda| \sim 0.4-0.5$\,\AA\ optically thin. On the contrary,
the line-of-sight ray at $\sim 0.9$\,Mm already samples layers with
high \hb\ wing emissivity above the temperature minimum region thus
producing an observable intensity contrast with the dark gap (top
panels of Figure~\ref{fig1}). We interpret the slight difference
between the height of the TM at 0.525\,Mm in the FAL-F model and the
height of the dark gap at 0.6\,Mm as a consequence of location of
electron density minimum in FAL-F at 0.6\,Mm. It shifts the opacity
minimum and the dark gap above the TM layer.

The gap near TM region was also detected in off-limb observations of
X8.2 class flare analyzed in \citet{Kuridzeetal2019,Kuridzeetal2020}
\citep[see Figure~2 in][]{Kuridzeetal2020}.  This indicates that
\hb\ off-limb observations can map the TM region under different
atmospheric conditions.

The dark gap is absent in the \hb\ line core images (bottom left panel
of Figure~\ref{fig3}). Its absence is due to a combination of opacity
maximum in the line core at all off-limb positions which is boosted by
high opaqueness of spicules and chromospheric canopy in the \hb\ core
hiding the dark gap beyond a thick opacity wall. The dark gap absence
in the \hb\ core is reproduced in the RH radiative transfer
calculations (bottom middle and right panels in Figure~\ref{fig3})
even without including a model of chromospheric canopy and spicules.

We demonstrate here that the off-limb emission in the \hb\ wings shows
high-contrast gap that maps the TM region.  Furthermore, our analyses
showed that enhanced microturbulence in the chromosphere is required
to reproduce the gap in the outer line wing spectral positions. The
nature of such an extra microturbulence can be various dynamical
phenomena, such as spicules, MHD waves, jets, shocks.  The work opens
a new window for studying the TM region by other lines of the Balmer
series and with more realistic chromospheric model atmosphere which
includes various dynamic phenomenons relevant for the chromosphere.
Trial \ha\ computations by the RH code and the FAL-C/F models (not
shown here) also show the dark gap in the wings. However, \ha\ has
larger optical thickness than \hb\ with narrower line width which
makes an effect of spicules on the emergent intensities more
important. These factors can make the \hb\ line a better candidate for
off-limb mapping of the TM region. The higher Balmer lines (e.g., \hg)
can likely produce even better contrast between the TM layer and the
upper layers.

\begin{acknowledgments}

D.K acknowledge Science and Technology Facilities Council
(STFC) grant ST/W000865/1, Leverhulme grant RPG2019-361 to Aberystwyth University and the
excellent facilities and support of SuperComputing Wales.  
The Swedish 1-m Solar Telescope is operated on the island of La Palma by the
Institute for Solar Physics of Stockholm University in the Spanish
Observatorio del Roque de los Muchachos of the Instituto de
Astrof\'{\i}sica de Canarias.  J.K. acknowledges the project VEGA
2/0048/20.  P.H. was supported by grant of the Czech Funding Agency
No. 19-09489S and by the program 'Excellence Initiative - Research
University' No. BPIDUB.4610.96.2021.KG of the University of Wroclaw.
R.O. acknowledges support from R+D+i project PID2020-112791GB-I00, 
financed by MCIN/AEI/10.13039/501100011033.  ChroTel is operated by the
  Leibniz-Institute for Solar Physics in Freiburg, Germany, at the
  Spanish Observatorio del Teide, Tenerife, Canary Islands. The
  ChroTel filtergraph has been developed by the Leibniz-Institute for
  Solar Physics in cooperation with the High Altitude Observatory in
  Boulder, CO, USA.
We would like to thank the anonymous referee for comments and suggestions that helped improve this paper.

\end{acknowledgments}

\software{RH code \citep{Uitenbroek2001}}
\facility{SST(CRISP, CHROMIS), Chromospheric Telescope (\href{http://oldwww.leibniz-kis.de/457/?L=1}{ChroTel})}

\bibliography{dkuridze_etal_2021}{}
\bibliographystyle{aasjournal}

\end{document}